\documentclass[3p,times,twocolumn]{elsarticle}

\usepackage{ecrc}

\volume{00}

\firstpage{1}

\journalname{Nuclear Physics B Proceedings Supplement}

\runauth{}

\jid{nuphbp}

\jnltitlelogo{Nuclear Physics B Proceedings Supplement}

\usepackage{amssymb}

\usepackage[figuresright]{rotating}

\def\CP{$ C \! P$ }

\def\ra{\rightarrow}

\def\Lbabar{\mbox{{\LARGE\sl B}\hspace{-0.15em}{\Large\sl A}\hspace{-0.07em}{\LARGE\sl B}\hspace{-0.15em}{\Large\sl A\hspace{-0.02em}R}}}

\def\sbabar{\mbox{{\normalsize \sl B}\hspace{-0.4em} {\small \sl A}\hspace{-0.03em}{\normalsize \sl B}\hspace{-0.4em} {\small \sl A\hspace{-0.02em}R}}}

\newcommand\T{\rule{0pt}{2.6ex}}       
\newcommand\B{\rule[-1.2ex]{0pt}{0pt}} 

\begin{document}

\begin{frontmatter}



\dochead{}

\title{Branching Fraction and \CP Asymmetry Measurements in Inclusive  $B \ra X_s \ell^+ \ell^- $ and $B \ra X_s \gamma $ Decays from \Lbabar}


\author{G. Eigen \\
representing the \sbabar\ collaboration}

\address{ Department of Physics, University of Bergen,
Allegaten 55, N-5007 Bergen, Norway\\
}

\begin{abstract}
We present an update on total and partial branching fractions and on \CP asymmetries in the semi-inclusive decay $B \ra X_s \ell^+ \ell^-$. Further, we summarize our results  on branching fractions and \CP asymmetries for semi-inclusive and fully-inclusive $B \ra X_s \gamma$ decays. We present the first result on the \CP asymmetry difference of charged and neutral $B \ra X_s \gamma$ decays yielding the first constraint on the ratio of Wilson coefficients $Im (C^{\rm eff}_8/C^{\rm eff}_7)$.
\end{abstract}


\end{frontmatter}


\section{Introduction}
\label{lab:introduction}

\begin{figure}[h]
\centering
\vskip -0.4cm
\includegraphics[width=0.6\columnwidth]{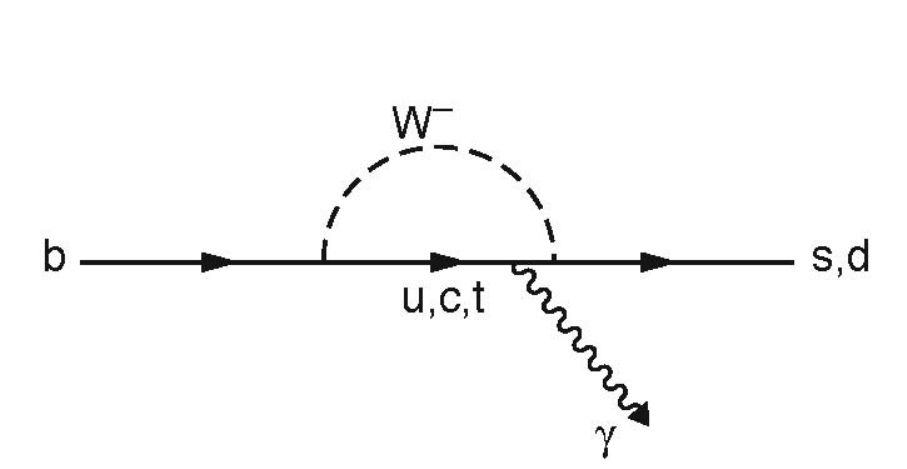}
\includegraphics[width=1.05\columnwidth]{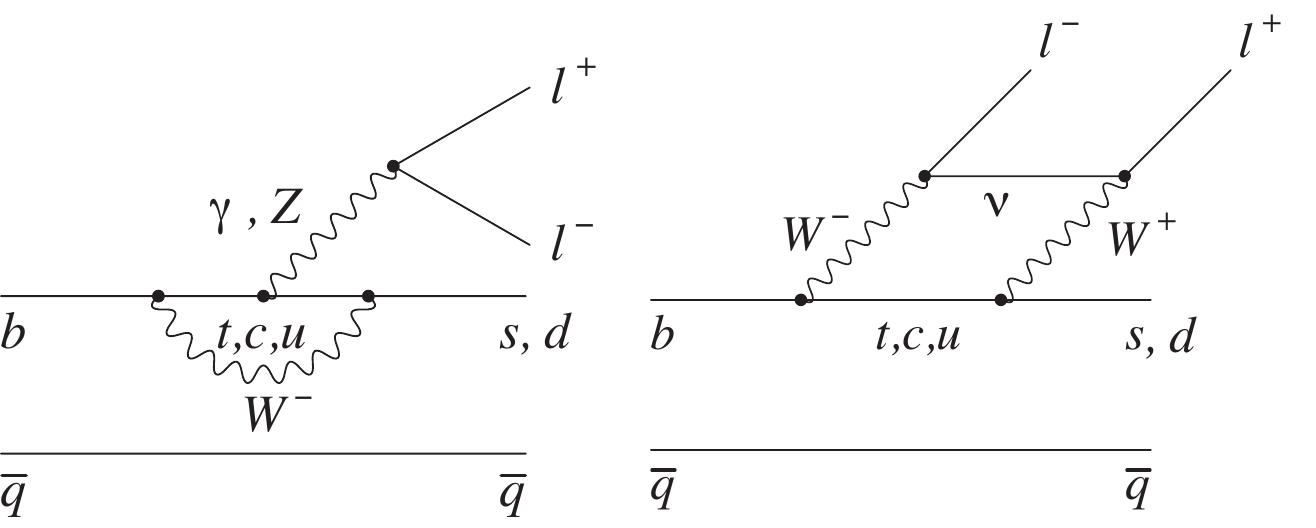}
\caption{Lowest-order diagrams for  $B \ra X_{s,d} \gamma$ (top) and $B \ra X_{s,d} \ell^+ \ell^-$ (bottom).}
 \label{fig:penguin}
\end{figure}

The decays $B \ra X_{s,d} \gamma$ and $ B \ra X_{s,d} \ell^+ \ell^-$ are flavor-changing neutral current (FCNC) processes that are forbidden in the Standard Model (SM) at tree level. However, they can proceed via penguin loops and box diagrams. Figure~\ref{fig:penguin} shows the lowest-order  diagrams for both processes. The effective Hamiltonian factorizes short-distance effects represented by perturbatively-calculable Wilson coefficients $(C_i)$~\cite{Wilson:1969zs,Wilson:1973jj} from long-distance effects specified by four-quark operators $({\cal O}_i)$:
\begin{equation}
H_{\rm eff} =\frac{G_F}{4 \pi} \Sigma_i V_{\rm xb}^* V_{\rm xs,d} C_i(\mu) {\cal O}_i.
\end{equation}

\begin{figure}[h]
\centering
\includegraphics[width=1.05\columnwidth]{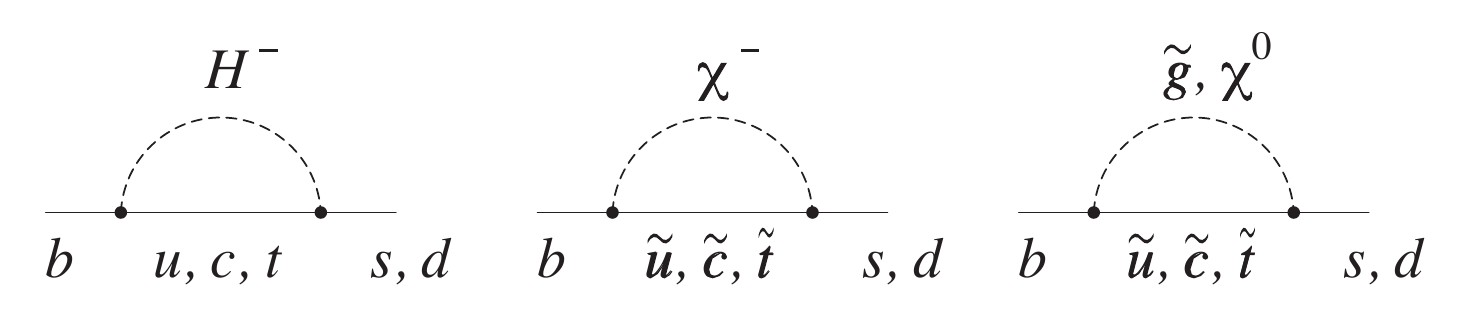}
\caption{New physics processes with a charged Higgs bosons (left), a chargino plus up-type squarks (middle) and neutralinos plus down-type squarks (right).}
 \label{fig:np}
\end{figure}

\noindent
Here, $G_F$ is the Fermi constant, $V_{\rm xb}^*$ and $V_{\rm xs,d}$ are CKM elements ($x=u, ~c, ~t$)  and $\mu$ is the renormalization scale. The operators have to be calculated using non-perturbative methods, such as the heavy quark expansion~\cite{Isgur:1988gb, Isgur:1989vq, Georgi:1990um, Grinstein:2004vb}. In $B \ra X_s \gamma$, the dominant contribution arises from the magnetic dipole operator ${\cal O}_7$ with a top quark in the loop. Thus, the branching fraction depends on the  Wilson coefficient $C^{\rm eff}_7 = -0.304$ (NNLL)~\cite{Buchalla:1995vs, Altmannshofer:2008dz}. Via operator mixing, the color-magnetic dipole operator ${\cal O}_8$ contributes in higher order with $C^{\rm eff}_8=-0.167$ (NNLL)~\cite{Buchalla:1995vs, Altmannshofer:2008dz}. In $B \ra X_s \ell^+ \ell^-$, the weak penguin and box diagrams contribute in addition. The vector part is represented by operator ${\cal O}_9$ with Wilson coefficient  $C^{\rm eff}_9=4.211 $~(NNLL)~\cite{Buchalla:1995vs, Altmannshofer:2008dz} while the axial-vector part is specified by operator ${\cal O}_{10}$ with Wilson coefficient $C^{\rm eff}_{10}=-4.103$~ (NNLL)~\cite{Buchalla:1995vs, Altmannshofer:2008dz}. Again, the top quark in the loop yields the most dominant contribution. New physics adds penguin and box diagrams with new particles modifying the  SM values of the Wilson coefficients. In addition, scalar and pseudoscalar couplings may contribute introducing new Wilson coefficients $C_S$ and $C_P$. Figure~\ref{fig:np} shows examples of new physics processes involving a charged Higgs, a chargino and neutralinos~\cite{Ali:2002jg, Lee:2008xc, Soni:2010xh, Oh:2011nb,  DescotesGenon:2011yn, Altmannshofer:2011gn, Kosnik:2012dj}. These rare decays probe new physics at a scale of a few TeV.

\section{Study of $B \ra X_s \ell^+ \ell^-$}

Using a semi-inclusive approach, we have updated the partial and total branching fraction measurements of $B \ra X_s \ell^+ \ell^-$ modes with the full \sbabar\ data sample of $471\times 10^6~ B \bar B$ events. We also perform the first measurement of direct \CP asymmetry. For measuring partial and total branching fractions,
we reconstruct  20 exclusive final states listed in Table~\ref{tab:xsll1}. After accounting for $K^0_L$ modes, $K^0_S \ra \pi^0\pi^0$ and $\pi^0$ Dalitz decays, they represent $70\%$ of the inclusive rate for hadronic masses $m_{X_s} < 1.8 \rm ~GeV$. Using JETSET fragmentation and theory predictions,  we extrapolate for the missing modes and those with $m_{X_s} > 1.8 ~\rm GeV$. We impose requirements on the beam-energy-substituted mass $m_{ES} = \sqrt{E^2_{CM} -p^{*2}_B}>5.225~\rm  GeV$ and on the energy difference $-0.1~ (0.05)<\Delta E=E^*_B -E_{CM}/2  <0.05$  for $X_s ~ e^+e^-~(X_s \mu \mu) $ modes where $E^*_B$ and $p^*_B$ are  $B$ momentum and $B$ energy in the center-of-mass (CM) frame and $E_{CM}$ is the total CM energy. We use no tagging of the $\bar B$ decay.

\begin{table}[!th]
\begin{center}
\caption{Exclusive modes used in the semi-inclusive $B \ra X_s \ell^+ \ell^-$ analysis. }
\vskip 0.2 cm
\begin{tabular}{|l||l|}  \hline\hline
Mode  & Mode      \\ \hline
$B^0 \ra K^0_S \mu^+ \mu^- $  & $B^+ \ra K^+ \mu^+ \mu^- $ \B \T \\
$B^0 \ra K^0_S e^+ e^- $  & $B^+ \ra K^+ e^+ e^- $ \B \T  \\
$B^0 \ra K^{*0} (K^0_S \pi^0) \mu^+ \mu^- $  & $B^+ \ra K^{*+} (K^+ \pi^0) \mu^+ \mu^- $ \B \T  \\
$B^0 \ra K^{*0} (K^+ \pi^-) \mu^+ \mu^- $  & $B^+ \ra K^{*+} (K^0_S \pi^+) \mu^+ \mu^- $  \B \T \\
$B^0 \ra K^{*0} (K^0_S \pi^0) e^+ e^- $  & $B^+ \ra K^{*+} (K^+ \pi^0) e^+ e^- $  \B \T  \\
$B^0 \ra K^{*0} (K^+ \pi^-) e^+ e^- $  & $B^+ \ra K^{*+} (K^0_S \pi^+) e^+ e- $  \B \T  \\
$B^0 \ra K^0_S \pi^+ \pi^-) \mu^+ \mu^- $  & $B^+ \ra K^0_S \pi^+ \pi^0 \mu^+ \mu^- $ \B \T  \\
$B^0 \ra K^+ \pi^- \pi^0 \mu^+ \mu^- $  & $B^+ \ra K^+ \pi^+ \pi^- \mu^+ \mu^- $ \B \T  \\
$B^0 \ra K^0_S \pi^+ \pi^-) e^+ e^- $  & $B^+ \ra K^0_S \pi^+ \pi^0 e^+ e^- $ \B \T  \\
$B^0 \ra K^+ \pi^- \pi^0 e^+ e^- $  & $B^+ \ra K^+ \pi^+ \pi^- e^+ e^- $ \B \T  \\
 \hline\hline
\end{tabular}
\label{tab:xsll1}
\end{center}
\end{table}

\begin{table}[!th]
\begin{center}
\caption{Definition of the $q^2$ bins. }
\vskip 0.2 cm
\begin{tabular}{|l|c|c|c|c|}  \hline\hline
$q^2$ bin & $q^2$  range $\rm [GeV^2/c^4]$ &  $m_{\ell \ell}$ range $\rm [GeV/c^2]$ & $m_{X_s}$ bin & $m_{X_s}$ range $ [\rm GeV/c^2] $   \B \T  \\ \hline
0 & 1.0 $< q^2 <$ 6.0 & 1.00 $< m_{\ell \ell}< $ 2.45 & &  \B \T  \\ \hline
1 & 0.1$< q^2 <$2.0 & 0.32$< m_{\ell \ell} < $1.41  & 1 & 0.4 -- 0.6 \B \T  \\
2 & 2.0$< q^2 <$4.3 & 1.41$< m_{\ell \ell} <$2.07   & 2 & 0.6 --1.0\B \T  \\
3 & 4.3$< q^2 <$8.1 & 2.07 $< m_{\ell \ell} <$2.6 & 3 & 1.0 --1.4\B \T  \\
4 & 10.1 $< q^2 <$12.9 & 3.18$< m_{\ell \ell} <$3.59 & 4 & 1.4 -- 1.8  \B \T  \\
5 & 14.2 $< q^2 <$ $ (m_B - m^*_K)^2$ & 3.77 $< m_{\ell \ell} <$ $ (m_B - m^*_K)$ & & \B \T  \\
 \hline\hline
\end{tabular}
\label{tab:xsll2}
\end{center}
\end{table}

\begin{figure}[!ht]
\begin{center}
\includegraphics[width=1.1\columnwidth]{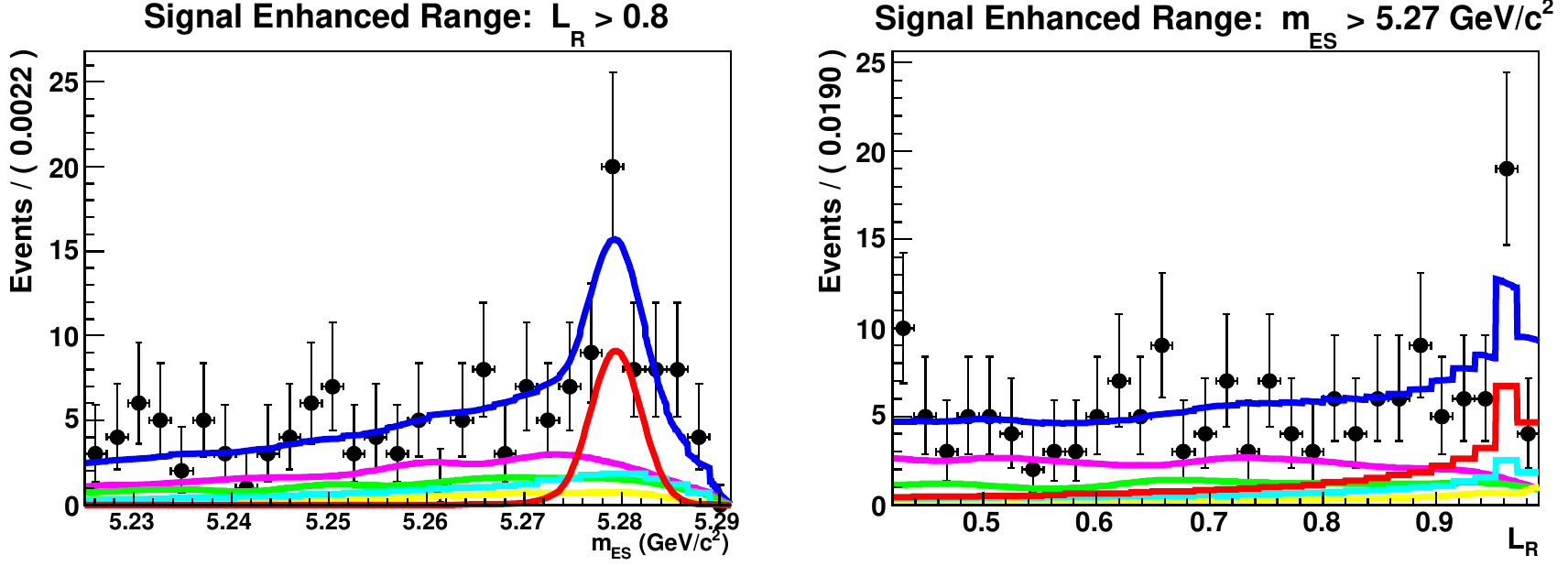}
\caption{Distributions of  $m_{ES}$  (left) and likelihood ratio (right) for $B \ra X_s e^+ e^-$ in $q^2$ bin $q_5$ showing data (points with error bars), the total fit (thick solid blue curves), signal component (red peaking curves),  signal cross feed (cyan/lgrey curves), $B \bar B $ background (magenta/dark grey smooth curve), $e^+ e^- \ra  q \bar q$ background (green/grey curves) and charmonium background (yellow/light grey curves).}
\label{fig:xseebf}
\end{center}
\end{figure}

\begin{figure}[!ht]
\begin{center}
\includegraphics[width=1.1\columnwidth]{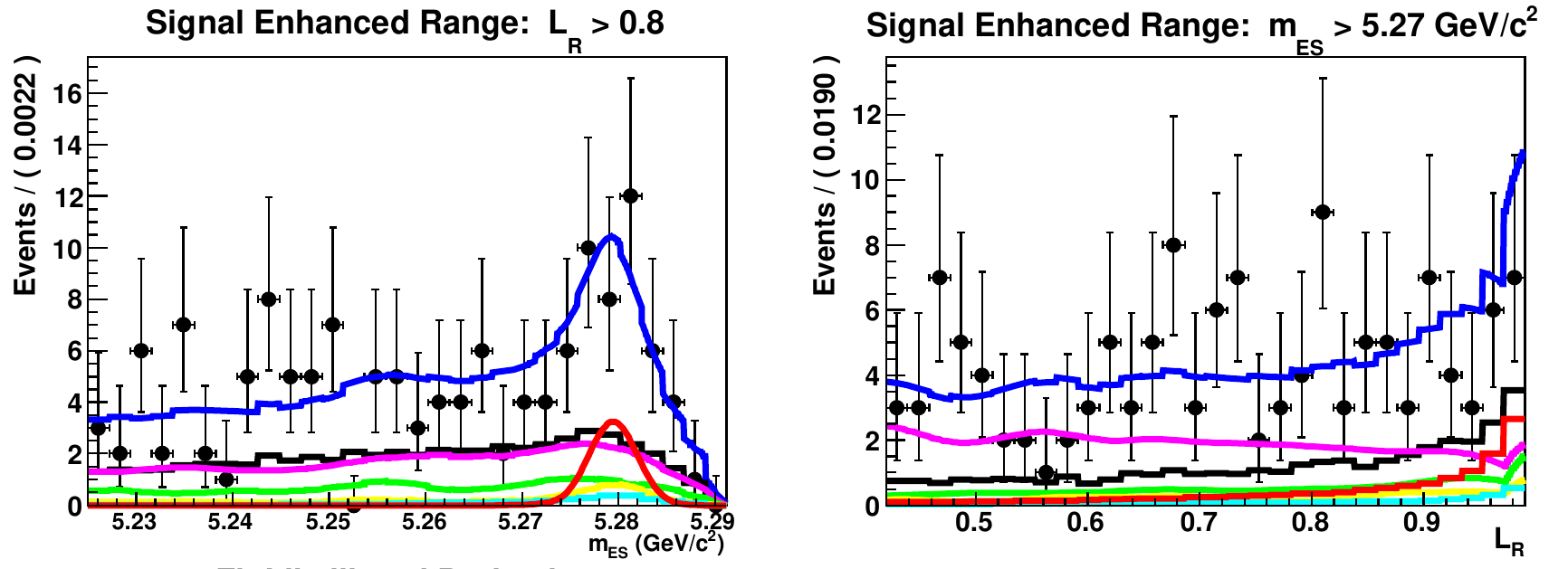}
\caption{Distributions of  $m_{ES}$  (left) and likelihood ratio (right) for $B \ra X_s \mu^+ \mu^-$ in $q^2$ bin $q_1$ showing data (points with error bars), the total fit (thick solid blue curves), signal component (red peaking curves),  signal cross feed (cyan/lgrey curves), $B \bar B $ background (magenta/dark grey smooth curve), $e^+ e^- \ra  q \bar q$ background (green/grey curves) and charmonium background (yellow/light grey curves)}
\label{fig:xsmmbf}
\end{center}
\end{figure}

To suppress  $e^+ e^- \ra q \bar q$ ($ q = u, d, s, c$) events and $B \bar B$ combinatorial background, we define boosted decision trees (BDT) for each $q^2$ bin in $e^+ e^-$ and $\mu^+ \mu^-$ separately (see Table 2). From these BDTs, we determine a likelihood ratio ($L_R$) to separate signal from $q \bar q$ and $B \bar B$ backgrounds. We
veto $J/\psi$ and $\psi(2S)$ mass regions and use them as control samples. Figures~\ref{fig:xseebf} and ~\ref{fig:xsmmbf} show the $m_{ES}$ and $L_R$ distributions for $e^+e^-$ modes in bin $q_5$ and for $\mu^+ \mu^-$ modes in bin $q_1$, respectively.

\begin{figure}[!ht]
\begin{center}
\includegraphics[width=0.95\columnwidth]{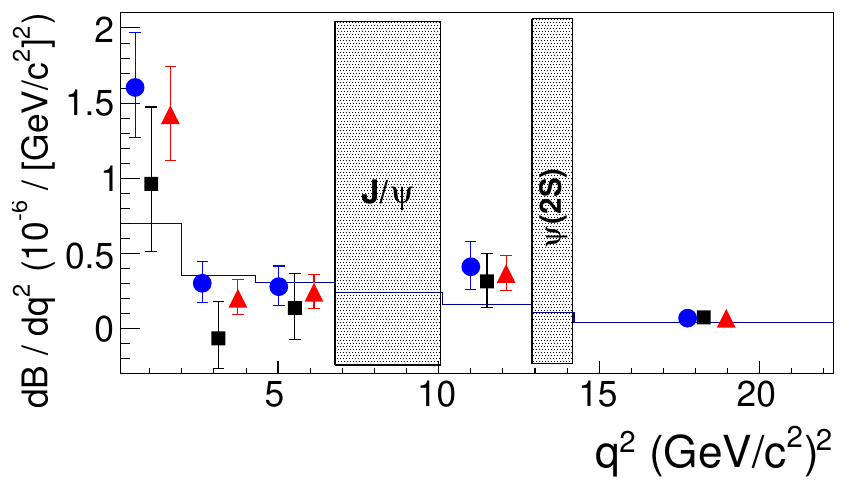}
\includegraphics[width=0.9\columnwidth]{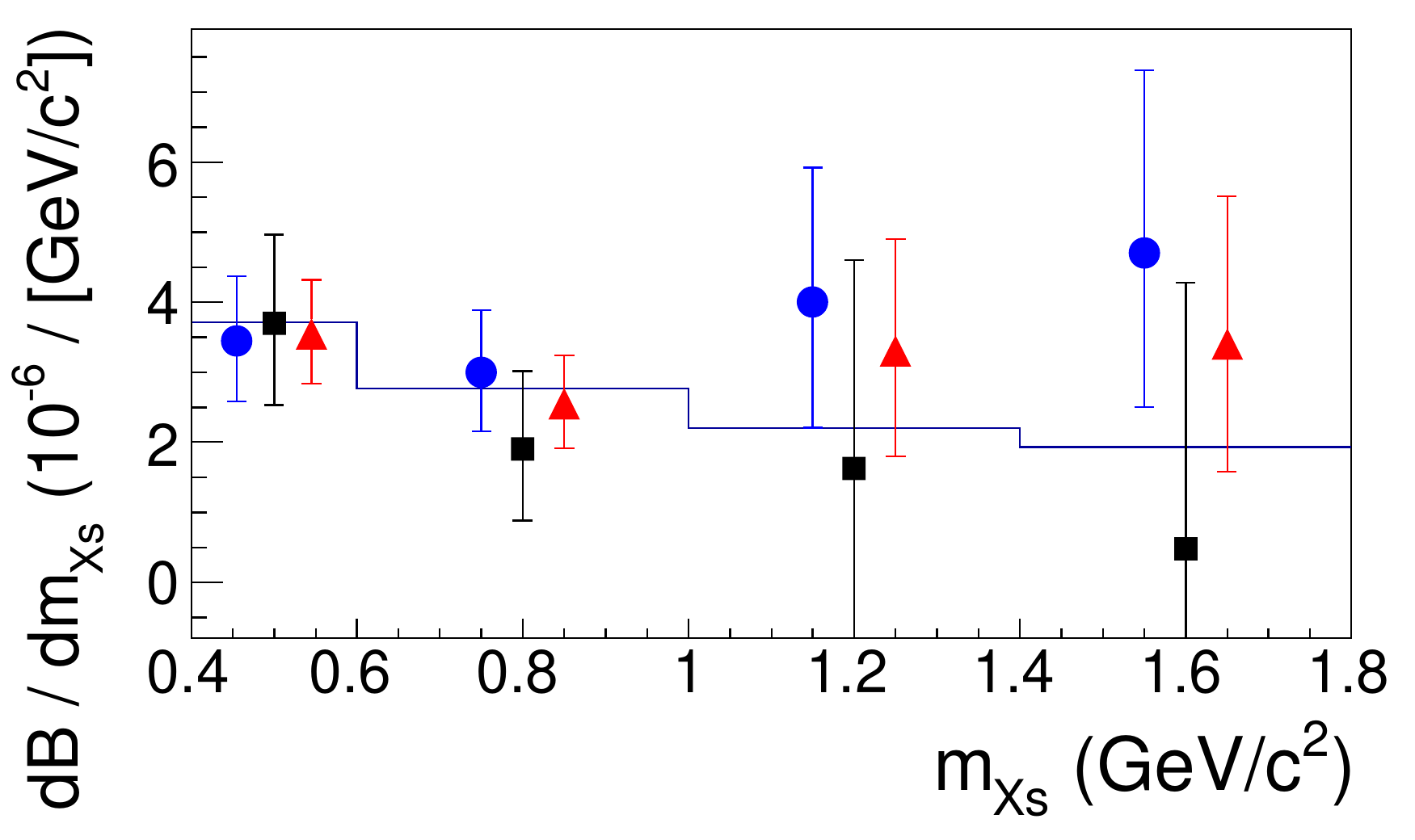}
\caption{Differential branching fraction of $B \ra X_s e^+ e^-$ (blue points), $B \ra X_s \mu^+ \mu^-$ (black squares), and
$B \ra X_s \ell^+ \ell^-$ (red triangles) versus $q^2$ (top) and versus $m_{X_s}$ (bottom) in comparison to the SM prediction (histogram). The grey-shaded bands show the $J/\psi$ and $\psi(2S)$ vetoed regions.}
\label{fig:xsllbf}
\end{center}
\end{figure}

We measure $d{\cal B}(B \ra X_s \ell^+ \ell^-)/dq^2$ in six bins of $q^2=m^2_{\ell \ell}$ and four bins of $m_{X_s}$ defined in Table~\ref{tab:xsll2}.  We extract the signal in each bin  from a two-dimensional fit to $m_{ES}$ and $L_R$.  Figure~\ref{fig:xsllbf} shows the differential branching faction as a function of $q^2$ (top) and $m_{X_s}$ (bottom)~\cite{Lees:2013nxa}. Table~\ref{tab:xsll3} summarizes the differential branching fractions in the low and high $q^2$ regions in comparison to the SM predictions~
\cite{Asatryan:2001zw, Asatryan:2002iy, Ghinculov:2002pe, Asatryan:2002iy, Gambino:2003zm, Ghinculov:2003bx, Bobeth:2003at, Ghinculov:2003qd,
Greub:2008cy, Huber:2007vv, Huber:2005ig, Beneke:2009az}.
In both regions of $q^2$, the differential branching fraction is in good agreement with the SM prediction. These results supersede the previous \sbabar\ measurements~\cite{Aubert:2004it} and are in good agreement with the Belle results~\cite{Iwasaki:2005sy}.

\begin{table}[!th]
\begin{center}
\caption{The $B \ra X_s \ell^+ \ell^-$ branching fraction measurements in the low and high $q^2$ regions~\cite{Lees:2013nxa} in comparison to the SM prediction.}
\vskip 0.2 cm
\begin{tabular}{|l|c|c|}  \hline\hline
Mode & \sbabar\  $[10^{-6}]$ & SM $[10^{-6}]$  \\ \hline
$q^2  [\rm GeV^2/c^4 ]$ & 1  -- 6   & 1 -- 6    \\ \hline
$B \ra X_s \mu^+ \mu^-$ & $0.66^{+0.82+0.30}_{-0.76-0.24}\pm 0.07 $ & $1.59\pm 0.11$  \B \T  \\
$B \ra X_s e^+ e^-$ & $1.93^{+0.47+0.21}_{-0.45-0.16}\pm 0.18 $ & $1.64\pm 0.11$   \B \T  \\
$B \ra X_s \ell^+ \ell^-$ & $1.60^{+0.41+0.17}_{-0.39-0.13}\pm 0.07$ & \B \T  \\ \hline
$q^2  [\rm GeV^2/c^4 ]$ & $ >14.2 $  & $ >14.2~  $ \B \T   \\ \hline
$B \ra X_s \mu^+ \mu^-$ &  $0.60^{+0.31+0.05}_{-0.29-0.04}\pm 0.00 $ &  $0.25^{+0.07}_{-0.06}$  \B \T   \\
$B \ra X_s e^+ e^-$ &  $0.56^{+0.19+0.03}_{-0.18-0.03}\pm 0.00$ & \B \T   \\
$B \ra X_s \ell^+ \ell^-$ &  $0.57^{+0.16+0.03}_{-0.15-0.02}\pm 0.00$ &    \B \T  \\
 \hline\hline
\end{tabular}
\label{tab:xsll3}
\end{center}
\end{table}

The direct \CP asymmetry is defined by:

\begin{equation}
{\cal A}_{C \! P}=\frac{{\cal B}(\bar B \ra \bar X_s \ell^+ \ell^-)-{\cal B}(B \ra  X_s \ell^+ \ell^-)}{{\cal B}(\bar B \ra \bar X_s \ell^+ \ell^-)+{\cal B}(B \ra  X_s \ell^+ \ell^-)}.
\end{equation}

\begin{figure}[!ht]
\begin{center}
\includegraphics[width=0.9\columnwidth] {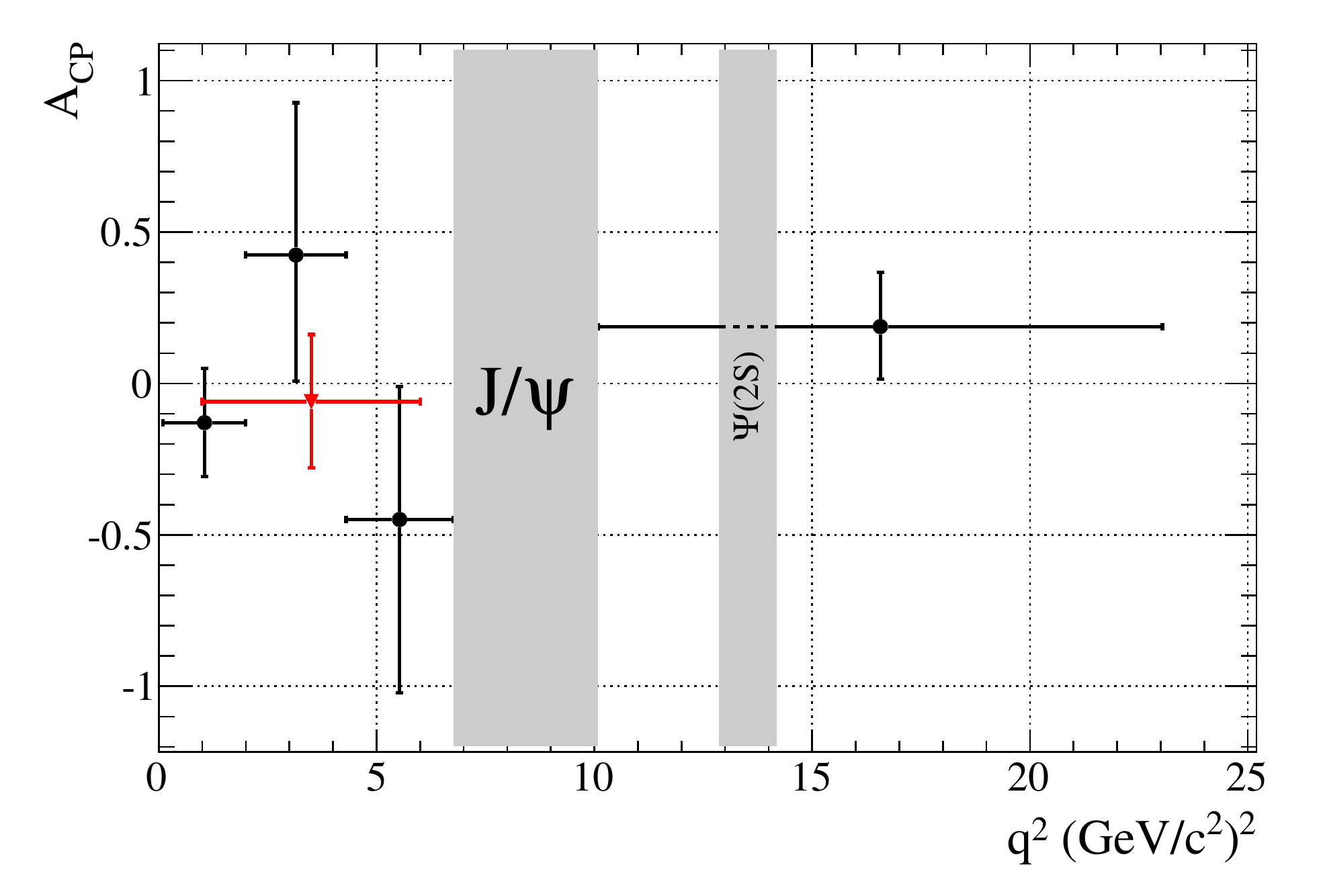}
\caption{The \CP asymmetry as a function of $q^2$. The grey-shaded bands show the $J/\psi$ and $\psi(2S)$ vetoed regions.}
\label{fig:xsllcp}
\end{center}
\end{figure}

We use 14 self-tagging modes consisting of all $B^+$ modes and the $B^0$ modes with decays to a $K^+$ listed in Table ~\ref{tab:xsll1} to measure ${\cal A}_{C \! P} (B \ra X_s  \ell^+ \ell^-)$ in five $q^2$ bins. Note that we have combined bins $q_4$ and $q_5$ due to low statistics. Figure~\ref{fig:xsllcp} shows the \CP asymmetry as a function of $q^2$. The SM prediction of the \CP asymmetry in the entire $q^2$ region is close to zero~\cite{Du:1995ez, Ali:1998sf, Bobeth:2008ij, Altmannshofer:2008dz}. In new physics models, however, ${\cal A}_{C \! P} $ may be significantly enhanced~\cite{Soni:2010xh, Alok:2008dj}. In the full range of $q^2$ we measure ${\cal A}_{C \! P}=0.04 \pm 0.11 \pm 0.01$~\cite{Lees:2013nxa}, which is in good agreement with the SM prediction. The \CP asymmetries in the five $q^2$ bins are also consistent with zero.

\section{Study of $ B \ra X_s \gamma$}

In the SM, the $B \ra X_s \gamma$ branching fraction is calculated in next-to-next leading order (4 loops) yielding
\begin{equation}
{\cal B}(B \ra X_s \gamma) =(3.15\pm 0.23) \times 10^{-4}
\end{equation}
for photon energies $E_\gamma > 1.6~\rm GeV$~\cite{Misiak:2006zs, Misiak:2006ab}. 

To extract the $B \ra X_s \gamma$ signal experimentally from $e^+ e^- \ra B \bar B$ and  $e^+ e^- \ra q \bar q$ backgrounds,  we use two very different strategies. The first strategy consists of a semi-inclusive approach in which we sum over  38 exclusive $B \ra X_s \gamma$ final states with $1K^\pm (\leq 1 K^0_S )$ or 3 $K^\pm$, $\leq 4 \pi (\leq 2 \pi^0)$,  and $\leq 1 \eta$. We use no tagging of the other $ B$ meson. We need to model the missing modes. Due to  large backgrounds, we select events with a minimum photon energy of $E_\gamma >1.9~\rm GeV$ and then extrapolate the branching fraction to photon energies  $E_\gamma >1.6~\rm GeV$.  With this approach, we measure the branching fraction, \CP asymmetry and the difference in \CP asymmetries between charged and neutral $B$ decays using $471 \times 10^6~ B \bar B$ events~\cite{Lees:2012wg}.

The second strategy is a fully inclusive approach. To suppress backgrounds from $B \bar B$ and $q \bar q$ decays, we impose stringent constraints on 
isolated photons to remove clusters that may have originated from $\pi^0$ and $\eta$ decays. We use a semileptonic  tag of the other $B$ meson and require a minimum photon energy of $E_\gamma >1.8 ~\rm GeV$ but impose no requirements on the hadronic mass system. Using $383 \times 10^6~ B \bar B$ events, we measure the $B \ra X_s \gamma $ branching fraction measurement  and the \CP asymmetry for $B \ra X_{s+d} \gamma$~\cite{Lees:2012ym,  Lees:2012ufa}.

\begin{figure}[h]
\centering
\includegraphics[width=1.0\columnwidth]{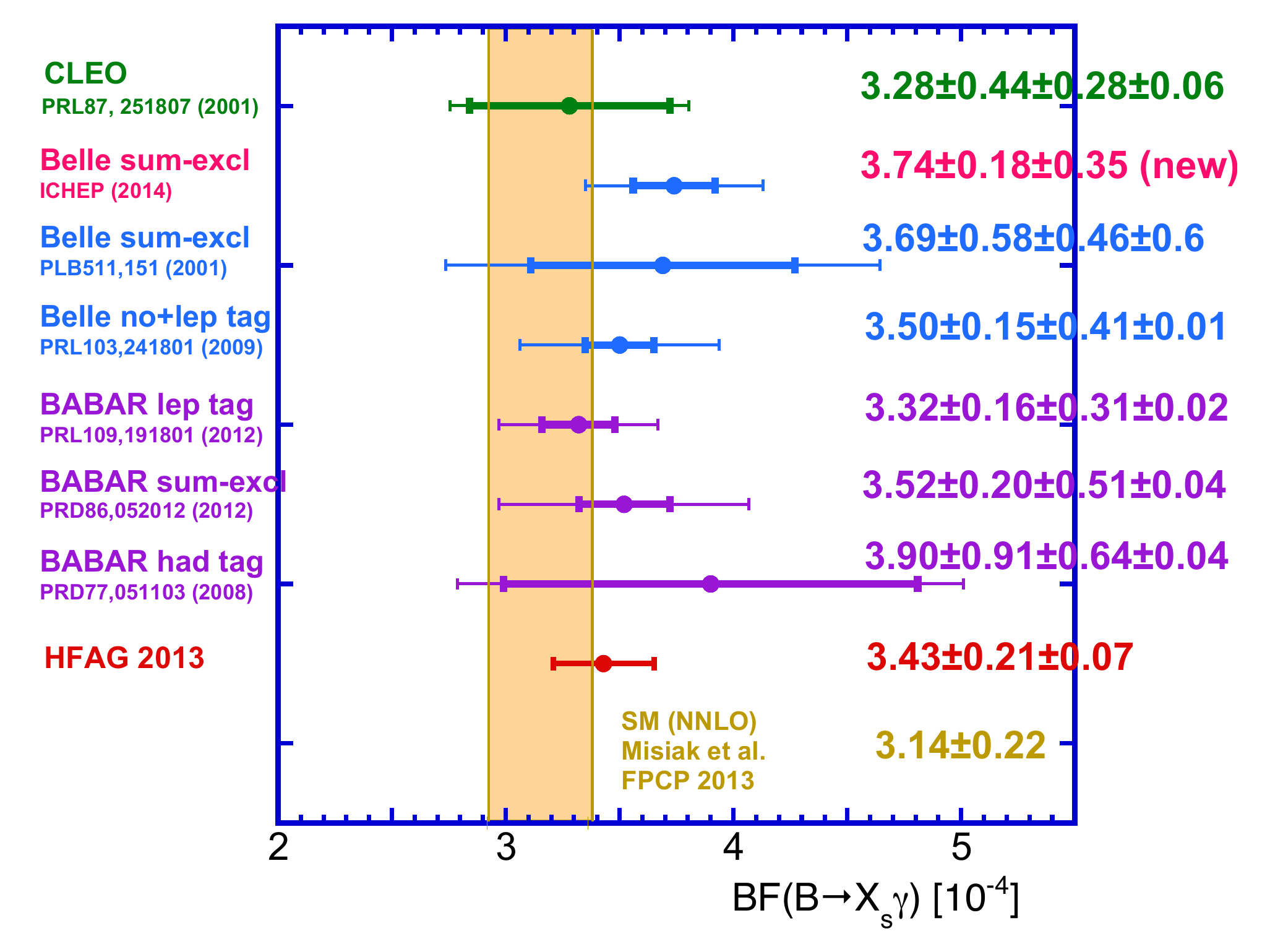}
\caption{Summary of ${\cal B}(B\ra X_s \gamma)$ measurements from BABAR~\cite{Lees:2012wg, Lees:2012ym, Lees:2012ufa, Aubert:2007my}, Belle~\cite{Abe:2001hk, Limosani:2009qg,  Sato:2014pjr},  CLEO~\cite{Chen:2001fja} and the HFAG average~\cite{Asner:2010qj} in comparison to the SM prediction~\cite{Misiak:2006zs, Misiak:2006ab} after extrapolation to $E^*_\gamma > 1.6 ~\rm GeV$. }
 \label{fig:bsgbf}
\end{figure}

Table~\ref{tab:bsgbf} summarizes our $B \ra X_s \gamma$ branching fraction measurements of the semi-inclusive  and  fully inclusive methods~\cite{Lees:2012wg, Lees:2012ym,  Lees:2012ufa}. Figure~\ref{fig:bsgbf} shows the \sbabar\ results extrapolated to a minimum photon energy of $1.6~\rm GeV$ in comparison to results from Belle~\cite{Abe:2001hk, Limosani:2009qg, Sato:2014pjr}, CLEO~\cite{Chen:2001fja} and the SM prediction~\cite{Misiak:2006zs, Misiak:2006ab}. Our results are in good agreement with those of the other experiments as well as the SM prediction.

\begin{table}[!th]
\begin{center}
\caption{ Our measurements of ${\cal B}(B \ra X_s \gamma)$ from the semi-inclusive~\cite{Lees:2012wg} and fully-inclusive~\cite{Lees:2012ym} analyses and their extrapolations to $E_\gamma > 1.6~\rm GeV$. The first uncertainty is statistical, the second is systematic and the third is from model  dependence and extrapolation to 1.6 GeV. }
\vskip 0.2 cm
\begin{tabular}{|l|c|c|}  \hline\hline
method & $E_\gamma >$ & ${\cal B}(B \ra X_s \gamma)~ [10^{-4}] $ \B \T \\ \hline
semi- & $1.9~\rm GeV$ & $3.29\pm 0.19 \pm 0.48\ \ \ \ \ \ \ \ \ \ \ $  \\
 exclusive & $1.6~\rm GeV$ & $3.52\pm 0.20 \pm 0.51 \pm 0.04$ \\ \hline
inclusive &  $1.8~\rm GeV$ &  $3.21\pm 0.15 \pm 0.29 \pm 0.08$  \\
 &  $1.6~\rm GeV$ &  $3.31\pm 0.16 \pm 0.30 \pm 0.10 $ \\
\hline\hline
\end{tabular}
\label{tab:bsgbf}
\end{center}
\end{table}

For the semi-inclusive method, the direct \CP asymmetry is defined by: 

\begin{equation}
{\cal A}_{C \! P}(X_s \gamma) =\frac{{\cal B}(\bar B \ra \bar X_s \gamma)-{\cal B}(B \ra  X_s \gamma)}{{\cal B}(\bar B \ra \bar X_s \gamma)+{\cal B}(B \ra  X_s \gamma)}.
\end{equation}

The SM prediction yields $ -0.6\%< {\cal A}_{C \! P}(B \ra X_s \gamma) < 2.8\%$~\cite{Kagan:1998bh, Benzke:2010tq}.
Using 16 self-tagging exclusive modes and $471 \times 10^6 ~B \bar B$ events, we measure  ${\cal A}_{C \! P} (B \ra X_s \gamma) =( 1.7 \pm 1.9_{stat} \pm 1.0_{sys})\%$~\cite{Lees:2014uoa}. This supersedes the old \sbabar\ measurement~\cite{Aubert:2008be}.We further measures the \CP asymmetry difference between charged and neutral $B$ decays:

\begin{equation}
\Delta {\cal A}_{C \! P}={\cal A}_{C \! P}(B^+ \ra X^+_s \gamma) - {\cal A}_{C \! P}(B^0 \ra X^0_s \gamma),
\end{equation}
which depends on the Wilson coefficients $C^{\rm eff}_7$ and $C^{\rm eff}_8$:

\begin{equation}
\hskip -0.8cm
\Delta {\cal A}_{C \! P} = 4 \pi^2 \alpha_s \frac{\bar \Lambda_{78}}{m_b} Im \frac{C^{\rm eff}_8}{C^{\rm eff}_7}
 \simeq 0.12  \frac{\bar \Lambda_{78}}{100~\rm MeV} Im \frac{C^{\rm eff}_8}{C^{\rm eff}_7}
\end{equation}
where the scale parameter $\bar \Lambda_{78}$ is constrained by $ 17~\rm MeV < \bar \Lambda_{78} < 190~ MeV$. In the SM,  $C^{\rm eff}_7$ and $C^{\rm eff}_8$
are real so that $\Delta {\cal A}_{C \! P}$ vanishes. However in new physics models, these Wilson coefficients may have imaginary parts yielding a non-vanishing $\Delta {\cal A}_{C \! P}$. 

From a  simultaneous fit to charged and neutral $B$ decays, we measure $\Delta {\cal A}_{C \! P}(B \ra X_s \gamma) = (5.0 \pm 3.9_{stat} \pm 1.5_{sys})\%$ from which we set an upper and lower limit at $90\% ~CL$ on $Im (C^{\rm eff}_8 / C^{\rm eff}_7)$~\cite{Lees:2014uoa}:
\begin{equation}
-1.64 < Im \frac{C^{\rm eff}_8}{C^{\rm eff}_7} <6.52~at~  90\%~ CL.
\end{equation}

\begin{figure}[h]
\centering
\includegraphics[width=0.8\columnwidth]{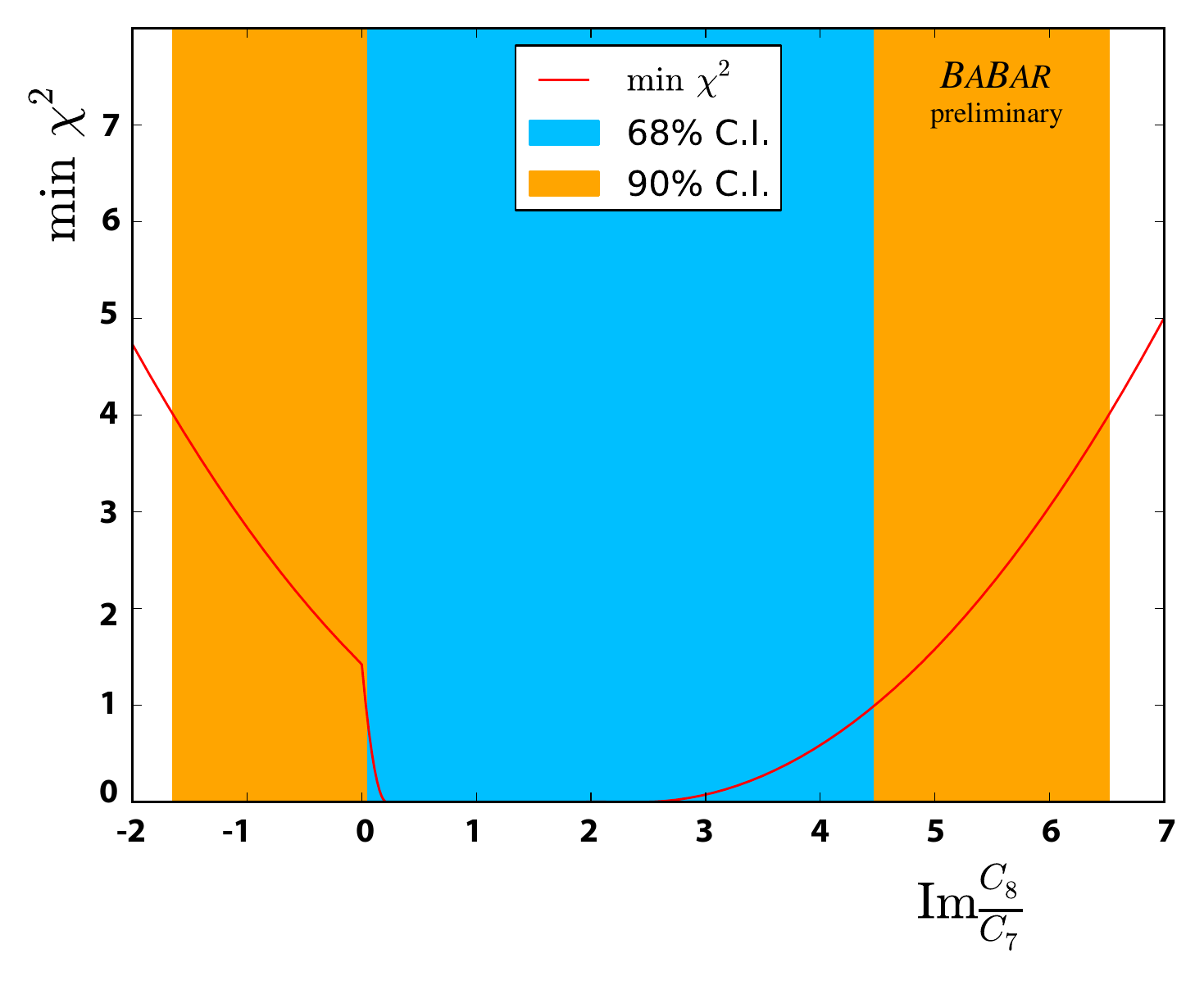}
\includegraphics[width=0.8\columnwidth]{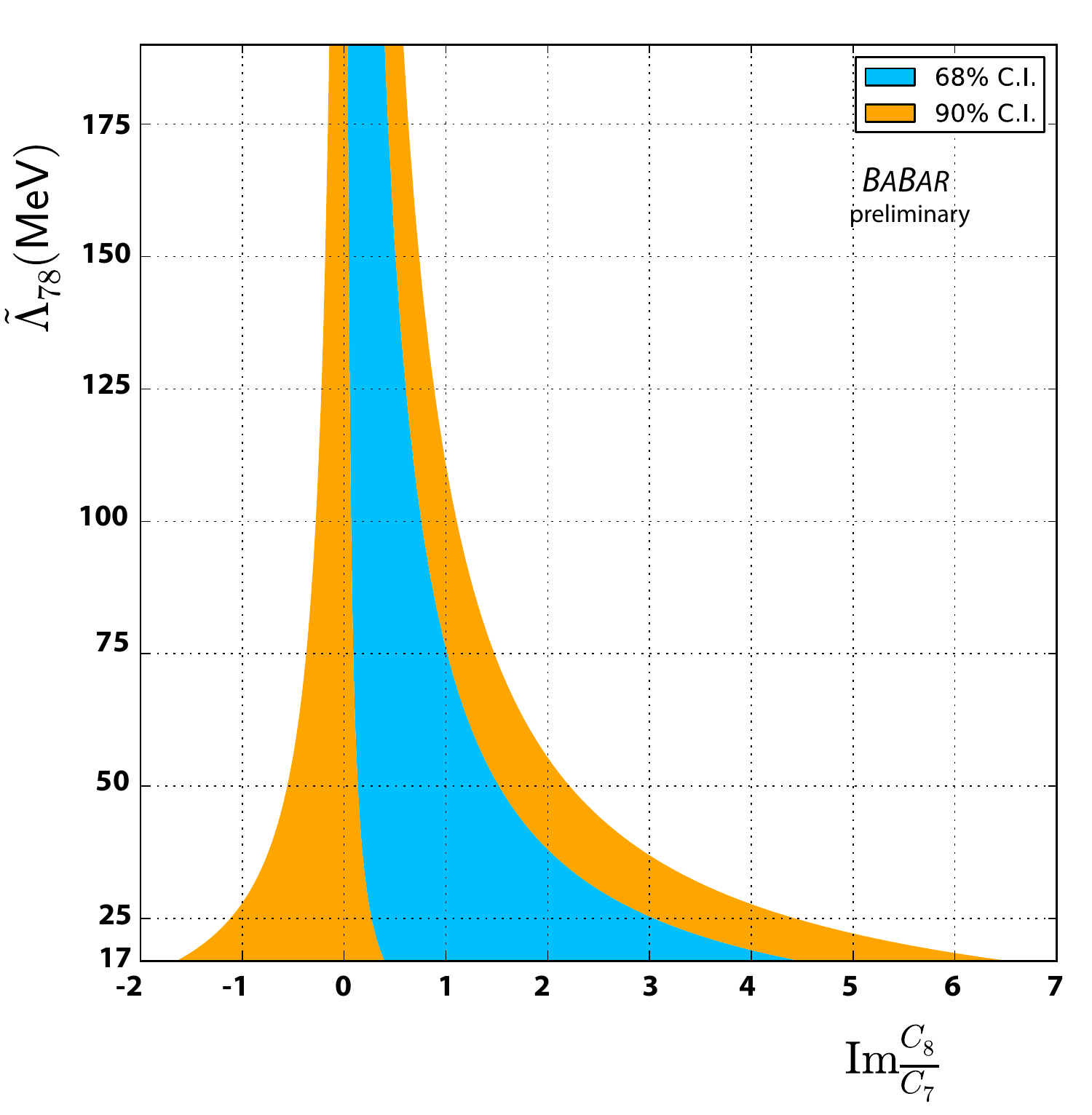}
\caption{The $\Delta \chi^2$ function versus ${\cal I}m(C^{\rm eff}_8/C^{\rm eff}_7)$ (top) and the dependence of $\bar \Lambda_{78}$ on ${\cal I}m(C^{\rm eff}_8/C^{\rm eff}_7)$ (bottom). The blue dark-shaded (orange light-shaded) regions show the $68\%~(90\%)$ CL intervals.}
 \label{fig:c78}
\end{figure}

This is the first $\Delta {\cal A}_{C\! P} $ measurements and the first constraint on $Im (C^{\rm eff}_8 / C^{\rm eff}_7) $. 
Figure~\ref{fig:c78} (top) shows the $\Delta \chi^2$ of the fit as a function of $Im (C^{\rm eff}_8 / C^{\rm eff}_7) $. The shape of $\Delta \chi^2$ as a function of  ${\cal I}m(C^{\rm eff}_8/C^{\rm eff}_7)$ is not parabolic indicating that the likelihood has a non-Gaussian shape. The reason is that $\Delta \chi^2$ is determined from all possible values of  $\bar \Lambda_{78}$. In the region $\sim 0.2 <  {\cal I}m(C^{\rm eff}_8/C^{\rm eff}_7) < \sim 2.6$ a change in 
${\cal I}m(C^{\rm eff}_8/C^{\rm eff}_7)$ $\Delta \chi^2$ can be compensated by a change in  $\bar \Lambda_{78}$ leaving 
 $\Delta \chi^2$ unchanged. For positive values larger (smaller) than 2.6 (0.2), $\Delta \chi^2$ increases slowly (rapidly), since  $\bar \Lambda_{78}$ remains nearly constant at the minimum value (increases rapidly). For negative ${\cal I}m(C^{\rm eff}_8/C^{\rm eff}_7)$ values, $\bar \Lambda_{78}$ starts to decrease again, which leads to a change in  the $\Delta \chi^2$ shape. Figure~\ref{fig:c78} (bottom) shows  $\bar \Lambda_{78}$ as a function of $Im (C^{\rm eff}_8 / C^{\rm eff}_7) $.

In the fully-inclusive analysis, the $B \ra X_d$ decay cannot be separated from the $B \ra X_s$ decay and we measure: 
\begin{equation}
\hskip -0.2cm
{\cal A}_{C \! P}(X_{s+d} \gamma) =\frac{{\cal B}(\bar B \ra \bar X_{s+d} \gamma)-{\cal B}(B \ra  X_{s+d} \gamma)}{{\cal B}(\bar B \ra \bar X_{s+d} \gamma)+{\cal B}(B \ra  X_{s+d} \gamma)}.
\end{equation}

\begin{figure}[h]
\centering
\includegraphics[width=1.1\columnwidth]{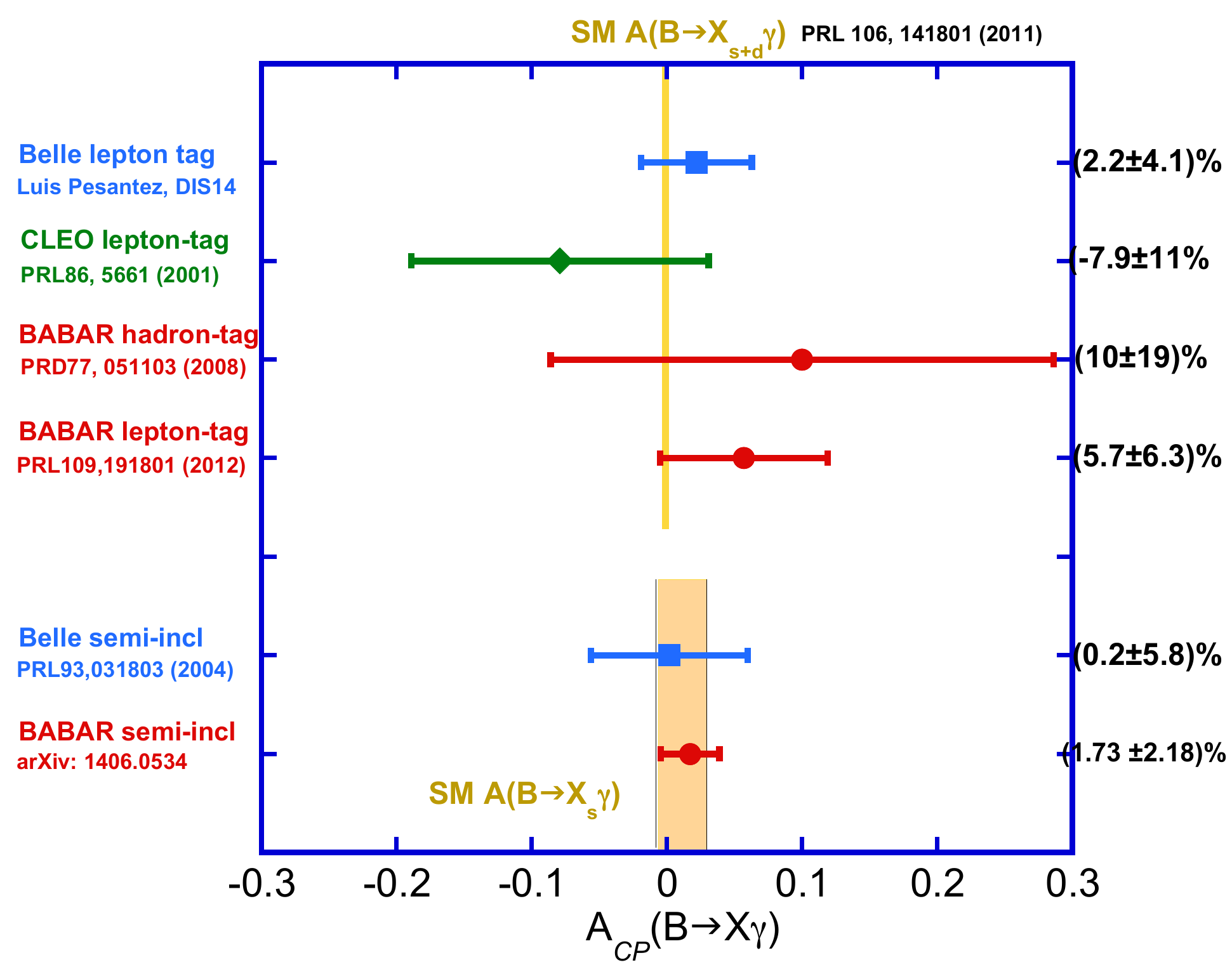}
\caption{Summary of  ${\cal A}_{C \! P}$ measurements for $B \ra X_s \gamma$ from semi-inclusive analyses~(\sbabar\ ~\cite{Lees:2014uoa}, Belle~\cite{Nishida:2003paa})  and for $B \ra X_{s+d} \gamma$ from fully inclusive analyses (\sbabar~\cite{Lees:2012ym, Lees:2012ufa, Aubert:2007my},  CLEO~\cite{Coan:2000pu}), Belle~\cite{Yook:2014kga} and the HFAG average~\cite{Asner:2010qj} in comparison to the SM prediction for $B \ra X_s \gamma$~\cite{Kagan:1998bh, Benzke:2010tq,  Hurth:2003dk}.}
 \label{fig:acpsg}
\end{figure}

In the SM, ${\cal A}_{C \! P}(B \ra X_{s+d} \gamma) $ is zero~\cite{Hurth:2003dk}. From the charge of the $B$ and $\bar  B$, we determine the \CP asymmetry. Using $383 \times 10^6$ $B \bar B$ events, we measure ${\cal A}_{C \! P}(B \ra X_{s+d} \gamma) = (5.7\pm 6.0 \pm 1.8) \%$, which is consistent with the SM prediction~\cite{Hurth:2003dk}. Figure~\ref{fig:acpsg} shows a summary of all \CP asymmetry measurements in comparison to the SM predictions.

\section{Conclusion}

We  performed the first ${\cal A}_{C \! P}$ measurement in five $q^2$ bins in semi-inclusive $ B \ra X_s \ell^+\ell^-$  decays and updated the differential branching fraction. The  $ B \ra X_s \ell^+\ell^-$ partial branching fractions and \CP asymmetries are in good
agreement with the SM predictions.  Our ${\cal A}_{C \! P}$ measurement in the semi-inclusive $B  \ra X_s \gamma$ decay is  the most precise \CP asymmetry measurement. The $ \Delta {\cal A}_{C \! P} (B \ra X_s \gamma)$ result yields first constraint on $Im (C^{ \rm e}_8/C^{ \rm e}_7)$. 
The  $B \ra X_s \gamma$ branching fractions and \CP asymmetries are both in good agreement with the SM predictions.
New progress on these inclusive decays will come from Belle II. For the $B \ra X_s \gamma$ and $B \ra X_s \ell^+ \ell^-$ semi-inclusive decays, we expect precision measurements. For the inclusive $B \ra X_s \gamma$  and $B \ra X_s \ell^+ \ell^-$ decays, we expect new possibilities by
tagging the other $\bar B$ meson via full $B$ reconstruction.

\section{Acknowledgments}

This work was supported by the Norwegian Research Council. I would like to thank members of the \sbabar\ collaboration for giving me the opportunity to present these results. In particular, I would like to thank Doug Roberts, Liang Sun and David Hitlin  for their fruitful suggestions.




\nocite{*}
\bibliographystyle{elsarticle-num}

\end{document}